\documentstyle[epsfig]{mn}

\newif\ifAMStwofonts

\ifoldfss
  \ifCUPmtlplainloaded \else
    \NewTextAlphabet{textbfit} {cmbxti10} {}
    \NewTextAlphabet{textbfss} {cmssbx10} {}
    \NewMathAlphabet{mathbfit} {cmbxti10} {} 
    \NewMathAlphabet{mathbfss} {cmssbx10} {} 
  \fi
  \ifAMStwofonts
    \ifCUPmtlplainloaded \else
      \NewSymbolFont{upmath} {eurm10}
      \NewSymbolFont{AMSa} {msam10}
      \NewMathSymbol{\upi}     {0}{upmath}{19}
      \NewMathSymbol{\umu}     {0}{upmath}{16}
      \NewMathSymbol{\upartial}{0}{upmath}{40}
      \NewMathSymbol{\leqslant}{3}{AMSa}{36}
      \NewMathSymbol{\geqslant}{3}{AMSa}{3E}

       \let\le=\leqslant
       \let\ge=\geqslant
    \fi
  \fi
\fi 

\ifnfssone
  \newmathalphabet{\mathit}
  \addtoversion{normal}{\mathit}{cmr}{m}{it}
  \addtoversion{bold}{\mathit}{cmr}{bx}{it}
  \newmathalphabet{\mathbfit} 
  \addtoversion{normal}{\mathbfit}{cmr}{bx}{it}
  \addtoversion{bold}{\mathbfit}{cmr}{bx}{it}
  \newmathalphabet{\mathbfss} 
  \addtoversion{normal}{\mathbfss}{cmss}{bx}{n}
  \addtoversion{bold}{\mathbfss}{cmss}{bx}{n}
  \ifAMStwofonts
    \ifCUPmtlplainloaded \else
      \UseAMStwoboldmath
      \makeatletter
      \new@mathgroup\upmath@group
      \define@mathgroup\mv@normal\upmath@group{eur}{m}{n}
      \define@mathgroup\mv@bold\upmath@group{eur}{b}{n}
      \edef\UPM{\hexnumber\upmath@group}
      \new@mathgroup\amsa@group
      \define@mathgroup\mv@normal\amsa@group{msa}{m}{n}
      \define@mathgroup\mv@bold\amsa@group{msa}{m}{n}
      \edef\AMSa{\hexnumber\amsa@group}
      \makeatother
      \mathchardef\upi="0\UPM19
      \mathchardef\umu="0\UPM16
      \mathchardef\upartial="0\UPM40
      \mathchardef\leqslant="3\AMSa36
      \mathchardef\geqslant="3\AMSa3E

       \let\le=\leqslant
       \let\ge=\geqslant
    \fi
  \fi
\fi 

\ifnfsstwo
  \DeclareMathAlphabet{\mathbfit}{OT1}{cmr}{bx}{it}
  \SetMathAlphabet\mathbfit{bold}{OT1}{cmr}{bx}{it}
  \DeclareMathAlphabet{\mathbfss}{OT1}{cmss}{bx}{n}
  \SetMathAlphabet\mathbfss{bold}{OT1}{cmss}{bx}{n}
  \ifAMStwofonts
    \ifCUPmtlplainloaded \else
      \DeclareSymbolFont{UPM}{U}{eur}{m}{n}
      \SetSymbolFont{UPM}{bold}{U}{eur}{b}{n}
      \DeclareSymbolFont{AMSa}{U}{msa}{m}{n}
      \DeclareMathSymbol{\upi}{0}{UPM}{"19}
      \DeclareMathSymbol{\umu}{0}{UPM}{"16}
      \DeclareMathSymbol{\upartial}{0}{UPM}{"40}
      \DeclareMathSymbol{\leqslant}{3}{AMSa}{"36}
      \DeclareMathSymbol{\geqslant}{3}{AMSa}{"3E}

       \let\le=\leqslant
       \let\ge=\geqslant
    \fi
  \fi
\fi 

\ifCUPmtlplainloaded \else
  \ifAMStwofonts \else 
    \def\upi{\pi}
    \def\umu{\mu}
    \def\upartial{\partial}
  \fi
\fi

\title{Cosmic evolution of metal densities: the enrichment of the Inter-Galactic Medium} 
\author[F. Calura, F. Matteucci]
       {F. Calura$^{1}$\thanks{E-mail: fcalura@ts.astro.it}, 
        F. Matteucci$^{1, 2}$\\
        (1) Dipartimento di Astronomia-Universit\'a di Trieste, Via G.
B. Tiepolo
	11, 34131 Trieste, Italy\\
	(2) INAF - Osservatorio Astronomico di Trieste, via G.B. Tiepolo 11,  
34131 Trieste, Italy \\
}
	
\date{Accepted for publication}

\pagerange{\pageref{firstpage}--\pageref{lastpage}}
\pubyear{2004}

\begin{document}

\maketitle

\label{firstpage}
\begin{abstract}
By means of chemo-photometric models for galaxies of different morhological, 
we have carried out a detailed study of the history of 
element production by spheroidal and dwarf irregular galaxies.  
Spheroidal galaxies suffer a strong and intense star formation episode at early times. 
In dwarf irregulars, the SFR proceeds  at a low regime but 
continuously.  Both galactic types enrich the IGM with metals, by means of galactic winds. 
We have assumed that the galaxy number density is fixed and normalized to the 
value of the optical luminosity function observed in the local universe. 
Our models allow us to investigate in detail how the metal 
fractions locked up in spheroid and dwarf irregular stars, in the ISM and ejected into the IGM have changed with cosmic time. 
By relaxing the instantaneous recycling approximation and taking into account stellar lifetimes, 
for the first time we have studied the evolution of the chemical abundance ratios in the IGM and compared 
our predictions 
with a set of observations by various authors. 
Our results indicate that 
the bulk of the IGM enrichment is due to spheroids, with dwarf irregular galaxies playing 
a negligible role. Our predictions grossly account for the [O/H] observed in the IGM at high redshift, but overestimate the [C/H]. 
Furthermore, it appears hard to reproduce the abundance ratios observed 
in the high-redshift IGM. Some possible explanations  are discussed in the text.  
This is the first attempt to study the abundance ratios in the IGM by means of detailed chemical evolution 
models which take into account the stellar lifetimes. Numerical simulations adopting our chemical evolution prescriptions could be 
useful to improve our understanding of the IGM chemical enrichment. 
\end{abstract} 


\section{Introduction} 
In the universe, metals can be found in different amounts in various 
environments. A fraction of all the metals produced so far is 
located in galaxies, locked up into living stars and stellar remnants, or contained in 
the inter-stellar gas. The remaining fraction of metals has been 
ejected by galaxies into the intergalactic and intracluster media (IGM, ICM). 
The amount of metals present in galaxies and IGM/ICM is a function of 
the cosmic epoch.
From the high-redshift universe to the present time, 
the observations indicate an evolution of the metallicity of galaxies (e.g Kobulnicky \& Kewley 2004).    
In lockstep with the growth of the galactic metallicity,   
the universe has experienced 
a progressive metal enrichment of the inter galactic gas.  
Metal abundances in the IGM are measurable by means of the quasar absorption lines  
(see Bechtold 2003, Pettini 2004). 
At high redshift, the observations report IGM metal abundances between $10^{-2.5}$ and 0.1 solar 
(Lu et al. 1998, Cowie \& Songaila 1998, Bergeron et al. 2002, Carswell 2004, Simcoe et al. 2002, Schaye et al. 2003, 
Simcoe et al. 2004a, Aguirre et al. 2004). In the local universe, the available data indicate 
an average IGM metallicity of the order of $\sim 0.1$ solar (Tripp et al. 2002, Shull et al. 2003).\\
Many theoretical studies have attempted to model the global chemical evolution of the universe and 
the chemical evolution of the IGM  
(Pei \& Fall 1995, 
Edmunds \& Phillips 1997, Malaney \& Chaboyer 1996, Pei, Fall \& 
Hauser 1999, Cen \& Ostriker 1999, Sadat Guiderdoni \& Silk 2001, Mathlin et al. 2001, De Lucia et al. 2003, 
Dunne et al. 2003), 
very fruitful in 
tracing the average properties of the universe but unsuited to 
study in detail the different contributions to the overall metal content of the universe from the 
single galactic types, namely spheroids, spiral discs and irregulars. 
Furthermore, many galactic evolution models compute metal production by adopting the instantaneous recycling approximation 
(IRA), which does not allow one to study the elements whose production occurs on
timescales not negligible relative to the age of the universe, such as C and Fe. 
In a recent paper, by means of chemical evolution models for galaxies of different morphological types, 
Calura \& Matteucci (2004, hereinafter CM04) have computed the total amount of metals produced by ellipticals, spirals and irregular galaxies 
and locked up in all the different components of the present-day universe.  \\
In the present paper, by means of the same detailed galactic chemical evolution models, 
we investigate how the metal 
fractions in the IGM have changed with cosmic time. 
One of the main assumptions of CM04 was that all galaxy types started forming at high redshift but with quite different 
star formation histories. In particular, elliptical galaxies are assumed to have formed in a very 
short timescale (less than 1 Gyr) and therefore, since they are the major metal producers, they are the most likely 
candidates to enrich the IGM at high redshift. A large cathegory of objects, experiencing very intense starbursts are 
observed at high redshift, thus confirming our assumption. 
These objects, which include the SCUBA (Blain et al. 1999, Trentham, Blain \& Goldader 1999)
and ultra luminous infrared galaxies (ULIRG, Rowan-Robinson 2000), very luminous in the infrared band but faint 
in the optical, dominate the cosmic star formation at redshift $z\sim 3$ (P\'erez-Gonz\'alez et al. 2005) and 
are natural candidates for 
massive proto-ellipticals (Ivison et al. 2000).\\
On the other hand, our assumptions are at variance with the hierarchical clustering scenario for galaxy formation. 
One of the main predictions of this scenario is that small structures, intended as dark matter halos, form first. 
As a consequence, the first galaxies to form are the smallest ones. The largest galaxies are formed later on  
by means of several merging episodes among gas rich galaxies, such as discs, occurring through the whole Hubble time 
(Kauffmann, White \& Guiderdoni 1993, Baugh et al. 1998, Cole et 
al. 2000, Somerville et al. 2001, Menci et al. 2002).  
These models were supported by some observational results like the apparent paucity of massive galaxies 
at $z \sim 1-2$ (Barger et al. 1999, Menanteau et al. 1999, Fontana et al. 1999, Drory et al. 2001, Dickinson et al. 2003). 
However, recent results seem to show that there is a consistent number of ellipticals at those redshifts 
(Schade et al. 1999, Im et al. 2002, Yamada et al. 2005). 
Beside spheroidal galaxies, we study also the contribution that dwarf irregular galaxies brought to the IGM metal enrichment. \\ 
We try to give a possible answer to several fundamental questions in galactic chemical evolution, such as:  
when was the IGM enriched with metals? 
Can massive spheroids forming  at high redshift account for the metals in the IGM? 
Is a ``normal'' (i.e. Salpeter-like) initial mass function (IMF) sufficient to enrich the IGM at the observed values? \\
This paper is organized as follows:   
in section 2 we recall all the main assumptions of the chemical evolution models and 
the method used to 
compute the comoving metal densities. In Section 3 we present the results and in Section 4 we draw the conclusions.

\section{The chemical evolution models}
We assume that the main contributors of the IGM enrichment are elliptical galaxies, the main spheroidal 
component of spiral galaxies, i.e. bulges, followed by dwarf irregular galaxies. 
Elliptical galaxies, spiral bulges and halos are all included in the cathegory of spheroids, and in this work 
are described by one single model. \\
In this paper, we assume that spiral discs do not develop galactic winds and do not contribute to the chemical enrichment of the IGM.  
This assumption is motivated by several reasons. 
First, in the disc of our own Galaxy, winds do not seem to have played an important role in its evolution. 
In fact, models of chemical evolution of the galactic disc with mass outflows do reproduce 
the abundances and abundance ratios observed in field stars (Tosi et al. 1998, Matteucci 2001). 
In general, the kinematical features of the interstellar gas in spiral discs seem rather different from the ones observed 
in local and distant objects experiencing winds and outflows. 
Zaritsky et al. (1990) have measured the velocities of  $H_{II}$ in local spirals, finding values of $\sigma_{V} \sim 10 km s^{-1}$. 
These values are  much lower than the ISM speeds in local starburst and Lyman-Break Galaxies, in which very energetic material moving at 
speeds $\ge 250  km s^{-1}$ has been detected (Pettini et al. 2002). 
The gas velocities measured in local spirals do not seem thus able to generate violent phenomena such as strong outflows or winds.\\
We calculate galaxy magnitudes by means of the spectro-photometric code by Jimenez et al. (1998)
This code allows us to compute self-consistently the 
photometric evolution of composite stellar populations taking into account the metallicity of the gas 
out of which the stars form at any time (see CM03 for a more detailed description).\\

\subsection{The spheroidal galaxy model}
The chemical evolution model representing spheroids (ellipticals and bulges) 
has been developed in order to reproduce a large set of observables  
concerning elliptical galaxies, such as abundance ratios and photometric properties (Matteucci 1994, Pipino \& Matteucci 2004). 
Detailed descriptions of the chemical evolution models used in this work 
can be found in Matteucci (1992, 1994). \\
In our picture, spheroids form as a result of the rapid collapse of a homogeneous sphere of
primordial gas where star formation is taking place at the same time as the collapse proceeds. 
Star formation is assumed to halt as the thermal energy $E_{th}$ of the ISM, heated by stellar winds and supernova (SN) explosions,
balances the binding energy of the gas. At this time a galactic wind occurs, sweeping away almost all of  
the residual gas. By means of the galactic wind, ellipticals enrich the IGM with metals.\\ 
The thermal energy $E_{th}$ of the ISM is calculated according to: 
\begin{equation} 
E_{th}=(E_{th})_{SN} + (E_{th})_{W} 
\end{equation} 
$(E_{th})_{SN}$ represents the energy injected by SNe and is given by: 
\begin{equation} 
(E_{th})_{SN} = \epsilon_{\rm SN Ia}E_0 R_{\rm SN Ia} 
+ \epsilon_{\rm SN II}E_0 R_{\rm SN II}\, 
\end{equation} 
where $E_0=10^{51}$erg is the energy released by each SN, 
$\epsilon_{\rm SN Ia}$ ($\epsilon_{\rm SN II}$) and $R_{\rm SN Ia}$ 
($R_{\rm SN II}$)
are the efficiency of energy transfer and the explosion rate for SNe Ia 
(SNe II), respectively. 
Our model takes into account also radiative cooling and follows the 
 prescriptions suggested by Cioffi, McKee \& Bertschinger (1988)  for the SN remnants evolution 
(see Pipino et al. 2002, 2005). According to this formulation,  
the net amount of energy transferred from SNe into the ISM is not larger than 20 \% of 
the initial blast wave energy. 
$(E_{th})_{W}$ indicates the energy injected by stellar winds and is given by:
\begin{equation}
(E_{th})_{W}=\int^{t}_{0}\int^{m_{up}}_{12}{\phi(m) \psi(t^{'})
 \eta_W E_{W} dmdt^{'}}
\end{equation}
Stellar winds give a negligible contribution to the total thermal 
energy of the gas. 
$E_{W} \sim 3 \cdot 10^{47}$ erg is the energy released into 
the ISM from a typical 
massive star ($\sim 20M_{\odot}$) through stellar winds during its 
lifetime and 
$\eta_W$ is the efficiency of energy transfer, which we assume to be $3\%$ according to Bradamante et al. (1998).\\
$\phi(m)$ represents the stellar initial mass function (IMF), which has been assumed to be constant in space and time. 
Unless otherwise stated, we assume a Salpeter IMF, which is given by:
\begin{equation}
\phi(m) = \phi_{0} \, m^{(-1+x)} 
\end{equation}

where $\phi_{0}$ is a normalization constant and $x=1.35$ over all the stellar mass range, $m_{inf} \le m/M_{\odot} \le m_{up}$, 
with $m_{inf}=0.1 M_{\odot}$ and  $m_{up}=100 M_{\odot}$. 
We will test also the possibility of an IMF flatter than the Salpeter one in the low-mass end, similar to the one 
found in the local disc of the MW (Chabrier 2003, Kroupa 2002).\\
$\psi(t)$ is the star formation rate (SFR), namely the fractional amount
of gas turning into stars per unit time. 
The SFR $\psi(t)$ (in $Gyr^{-1}$) is given by:

\begin{equation}
\psi(t) = \nu G(t) 
\end{equation}

The quantity $G(t)= M_{g}(t)/M_{tot}$ is the total fractional mass of gas
present in the galaxy at time t, being $M_{g}(t)$ and $M_{tot}(t)$ the gas mass  and total 
mass at the time $t$, respectively. 
The quantity $\nu$ is the efficiency of star formation, namely the inverse of
the time scale for star formation and for a typical elliptical with $10^{11} M_{\odot}$ of baryonic matter  
is assumed to be $\sim 10 $ Gyr$^{-1}$. 
As shown by Matteucci (1994), to reproduce the 
observed correlation between the magnesium index $Mg_{2}$ and the iron 
index $<Fe>$ in ellipticals, this quantity has to be higher in more massive galaxies. 
This leads to the so-called ``inverse wind'' scenario, 
in the sense that large galaxies develop a wind before the small ones. \\
The SFR is 
assumed to drop to zero at the onset of a galactic wind.\\
The binding energy $E_{B}(t)$ is given by the expression (see Pipino et al. 2002): 
\begin{equation}
E_{B}(t)=W_{L}(t) + W_{DM}(t) 
\end{equation} 
where $W_{L}(t)$ represents the gravitational energy of the gas due to the luminous matter:
\begin{equation}
W_{L}(t) = -\frac{0.5 G M_{g}(t) M_{L}(t)}{r_{L}}
\end{equation} 
where $M_{g}(t)$, $M_{L}(t)$ and $r_{L}$ represent, at the instant $t$, the gas mass, the total luminous mass and the effective 
radius, respectively.  
$W_{DM}(t)$ is the gravitational energy due to the dark matter (DM), and is given by: 
\begin{equation}
W_{DM}(t) = -\frac{G M_{g}(t) M_{Dark}(t)}{r_{L}}  W^{'}_{DM}
\end{equation}
$M_{Dark}(t)=10 M_{L}$ is the amount of DM present in the galaxy, $W^{'}_{DM}$ takes into account the distribution of the dark matter relative to the luminous one: 
\begin{equation}
W^{'}_{DM}(t) = \frac{1}{2 \pi} \frac{r_{L}}{r_{D}} \lbrack 1+1.37 ( r_{L}/r_{D} ) \rbrack  
\end{equation}
$r_{L}/r_{D}$ is the ratio between the effective radius and the core radius of the 
DM. We set $r_{L}/r_{D} =0.1$, in agreement with the results by 
Bertin et al. (1992). 

 The prescriptions for the DM halo adopted here, namely a rather heavy but diffuse DM halo, are supported by several 
observational results (Gerhard et al. 2001, Romanowsky et al. 2003, Samurovic \& Danziger 2005), which have shown that in several early 
type galaxies,  there is 
no strong evidence for dark haloes out to several effective radii.  
The adopted dark matter profile is hence different than a cuspy profile, like the one suggested 
by Navarro, Frenk \& White (NFW, 1996). 
Several recent theoretical results seem to exclude a cuspy DM profile or
presenting a peak in the center as the one by NFW (Mamon \& Lokas 2005, Merritt et al. 2005). 
It is worth to note that in SPH simulations of elliptical galaxies, galactic winds in 
very massive galaxies are difficult to explain owing to deep 
dynamical potential in the central regions (Kobayashi 2005). 
It is possible that in such simulations, the adoption of a more 
diffuse DM profile, like the one of this paper, could solve this problem. 
Concerning the baryonic matter, we take into
account the self gravity of the gas and we assume that the gas
distribution is the same as the one for the stars at any time. 
Let $G_{i}$ be the fractional mass of the element $i$ in the gas 
within a spheroidal galaxy, its temporal evolution is described by the basic equation:\\
\begin{equation}
\dot{G_{i}}=-\psi(t)X_{i}(t) + R_{i}(t) - 
(\dot{G_{i}})_{out}\\
\end{equation} 
If $M_{g}(t)$ is the mass of the gas at the time $t$, 
$G_{i}(t)=M_{g}(t)X_{i}(t)/M_{tot}$ is the gas mass in the form of an
element $i$ normalized to the total initial mass $M_{tot}$. 
If $G(t)=M_{g}(t)/M_{tot}$, 
the quantity $X_{i}(t)=G_{i}(t)/G(t)$ represents the abundance in mass of an element $i$, with 
the sum  over all the elements in the gas mixture being equal to unity. 
 $R_{i}(t)$ represents the returned
fraction of matter (both newly processed and already present in the star) in the form of an element $i$ that the stars eject into the ISM through stellar winds and 
SN explosions. This term contains all the prescriptions regarding the stellar yields and
the SN progenitor models, and is given by:\\

\begin{math}
  R_{i}(t)=
   \int_{M_{L}}^{M_{Bm}}\psi(t-\tau_m)
Q_{mi}(t-\tau_m)\phi(m)dm \nonumber    +
\end{math}
\begin{math}
 A\int_{M_{Bm}}^{M_{BM}}
\phi(m)\nonumber  
  \cdot[\int_{\mu_{min}}
^{0.5}f(\mu)\psi(t-\tau_{m2}) 
Q_{mi}(t-\tau_{m2})d\mu]dm\nonumber +
\end{math}
\begin{math}
  + (1-A)\int_{M_{Bm}}^
{M_{BM}}\psi(t-\tau_{m})Q_{mi}(t-\tau_m)\phi(m)dm\nonumber +
\end{math}
\begin{equation}
   \int_{M_{BM}}^{M_U}\psi(t-\tau_m)Q_{mi}(t-\tau_m) 
\phi(m)dm \nonumber
\end{equation}

Each one of these integrals represents the contribution of different stellar mass ranges, 
including type Ia SNe (second integral). 
In particular, $Q_{mi}(t-\tau_m)=Q_{mi} \, X(t-\tau_m)$, where $Q_{mi}$ is a matrix which calculates for any star 
of a given mass $m$ the amount of the newly processed and already present 
element $i$, which is returned to the ISM. 
The quantity $\tau_m$ is the lifetime of a star of mass $m$. The constant $A$ is the fraction of binary systems which produce type Ia SNe. 
For all the details of the calculations of the type Ia SN rate, we address the reader to Greggio \& Renzini (1983) and Matteucci \& Greggio 
(1986). 
The term $(\dot{G_{i}})_{out}$ accounts for the presence of gas outflows, occurring  
by means of SN-driven galactic winds. 
 $(\dot{G_{i}})_{out}$ is simply equal to the rate of mass 
loss from dying stars after the occurrence of the wind, since all the matter present in the ISM is lost 
on a timescale of a few $10^{8}$ years. 
We assume that metals ejected into the IGM by any galactic type do not accrete on the other galactic types. 
Recent studies have indicated that, if the star formation is rapid ($\le 0.5$ Gyr), the presence of an infall term has 
negligible effects on the predicted chemical abundances (Pipino \& Matteucci 2004). 
For this reason, we assume that all the gas is already present at the beginning of star formation, 
which is equivalent to a very short infall episode.\\

\subsection{The dwarf irregular galaxy model}
The dwarf irregular galaxies are assumed to 
assemble all of their gas by means of a continuous infall 
of pristine gas. 
Hence, the chemical evolution equation is slightly different than eq. 10, since it contains also an infall term, which is 
given by:
\begin{eqnarray}
 (\dot{G_{i}})_{inf} = \, C \, {(X_{i})_{inf}e^{-(t/\tau)} \over M_L} 
\end{eqnarray}
where $(X_{i})_{inf}$ is the abundance of the element {\it i} in the infalling
gas, assumed to be primordial, $\tau$ is the time scale of mass accretion 
and $C$ is a constant obtained by imposing
to reproduce the total luminous mass $M_L$ at the present time $t_{G}$.
The parameter $\tau$ has been assumed to be the same for all dwarf 
irregulars and short enough to avoid unlikely high infall rates at
the present time ($\tau=0.5\cdot 10^{9}$ years). \\
As in the case of ellipticals, dwarf irregulars are assumed to be embedded in a dark matter halo. 
The prescriptions for the dark matter halo are similar to those 
adopted for the ellipticals,
namely a massive but diffuse dark halo  
(see Bradamante et al. 1998 for details). 
The SFR proceeds  at a low regime but 
continuously.  
The SFR expression is identical to the one assumed for ellipticals (see eq. 5
), but the SF efficiency adopted in this case is $\nu=0.05 Gyr^{-1}$. 
The SF, if strong enough, can also trigger a galactic wind, in principle easy 
to develop in irregular galaxies owing to their shallow gravitational potential wells.  
The conditions for the onset of the wind  are identical to the ones described for elliptical galaxies 
(Bradamante et al. 1998, Recchi et al. 2002).\\
Also in this case, we assume a Salpeter IMF. 
The chemical evolution model for dwarf irregulars allows us to reproduce the main features of local blue compact dwarfs (BCD) and dwarf irregular galaxies. 
The predicted present-day ISM O abundance for the typical dwarf galaxy model is 12+log(O/H)= 8.5, with local observations of BCDs 
indicating values between 7.15  and 9  (Shi et al.  2005). 
The predicted HI mass to blue luminosity ratio is $M_{HI}/L_{B}=0.5 M_{\odot}/L_{\odot}$, 
with the observations indicating values of $0.2-2  M_{\odot}/L_{\odot}$ (Garland et al. 2004). 
The predicted SFR at the present time is $0.1  M_{\odot}/yr$, while local observation in irregular magellanic and BCDs show values 
of $0.01 - 1  M_{\odot}/yr$ (Hopkins et al. 2002, Hunter \& Elmegreen 2004).\\

\subsection{Stellar Yields}
In both the models used in this work, 
the chemical enrichment of the ISM occurs by means of three processes: stellar winds, type Ia and type II SN explosions. 
Low and intermediate ($0.8 \le M/M_{\odot} \le 8$)  mass stars enrich the ISM by means of quiescent mass loss, or stellar winds. 
For these kind of stars, we adopt the nucleosynthetic prescriptions by van den Hoeck \& Groenewegen (1997).\\
Type Ia SN are assumed to occur in binary systems with total masses $3 \le M_{B}/M_{\odot} \le 16$, 
formed by a white dwarf (WD) accreting mass from a 
secondary star until the 
Chandrasekhar mass ($\sim 1.4 M_{\odot}$) is reached (Greggio \& Renzini 1983). 
At this point, the star explodes by means of the thermonuclear runaway and no remnant is left. \\
Single massive (M $> 8 M_{\odot}$) stars enrich the ISM by means of type II SN explosions. 
After the explosion, a neutron star or a black hole can be left as remnants. 
For massive stars and type Ia SNe, we adopt the empirical yields suggested by  Fran\c cois et al. (2004),   
which are substantially based on the Woosley \& Weaver (1995) and Iwamoto et al. (1999) yields, respectively. 
For C production, we do take 
into account also rotation in massive stars, consistently  with Meynet 
and Maeder (2005). 
Therefore, the results for these elements are slightly different than the ones obtained in Calura \& Matteucci 2004.

\subsection{Comoving densities of chemical elements and IGM abundances}

To study galaxy evolution, we assume a pure luminosity evolution picture.  
According to this assumption, galaxies evolve 
only in luminosity and their number density is constant in time. 
Such a picture can account for many observables, such as the evolution 
of the galaxy luminosity density in various bands and the cosmic supernova rates (Calura \& Matteucci 2003, hereinafter CM03).\\
 The galactic morphological types considered in this work are two: spheroids and dwarf irregulars. 
The production rate $\dot{\rho}_{i}^{k}(z)$, 
for the $i-$th element and the galaxies of $k-$th morphological type as a function of the redshift $z$ is given by:\\
\begin{equation}
\dot{\rho}_{i}^{k}(M_{\odot} \, yr^{-1} \, Mpc^{-3}) (z)=\rho_{B}^{k}(z) (M/L)_{B}^{k}(z) \gamma_{i}^{k}(z) \\
\end{equation} 
where $\rho_{B}^{k}(z)$ is the B band luminosity density generated by galaxies of $k-$th morphological type and 
is the result of the integral over 
all luminosities of the morphological B-band luminosity function (LF) $\Phi^{k}(L_{B})(z)$, namely:\\
\begin{equation}
\rho_{B}^{k}(z)= \int{\Phi^{k}(L_{B})(z)\, (L_{B}/L^{\ast}_{B}(z))\, dL_{B}} 
\end{equation}
The general LF $\Phi(L_{B})$ is parametrized according to the form defined
 by Schechter (1976):
 \begin{equation}
 \Phi(L_{B})(z)=
 \Phi^{\ast}\,(L_{B}/L_{B}^{\ast}(z))^{-\alpha}\,exp(-L_{B}/L_{B}^{\ast}(z))
 \end{equation}
 where $\Phi^{\ast}$ is a normalization constant related to the number of luminous galaxies per unit
 volume, 
 $L_{B}^{\ast}(z)$ is a characteristic luminosity and $\alpha$ is the faint-end slope. \\
At redshift zero, we adopt the morphological LF as observed by Marzke et al. (1998). 
At redshift larger than zero, we have calculated the LF  
by applying the evolutionary corrections calculated by means of the spectrophotometric model 
by Jimenez et al. (1998) to the local LF. \\
The evolution of the stellar, gaseous and total mass is a direct prediction of the 
chemical evolution models, whereas the evolution of the B luminosity $L_{B}$ is calculated by means of the spectro-photometric 
code. 
$(M/L)_{B}^{k}(z)$ represents  the B mass-to-light ratio for $k-$th galaxies, where $M$ stands for the total baryonic mass. 
$\gamma_{i}^{k}(z)$ represents 
the rate of production of the $i-$th element by a stellar generation (only the newly processed fraction) in $k-$th galaxies. 
At a given time $t$, it is given by:\\
\begin{equation}
\gamma_{i}^{k}(t) =  \int_{m(t)}^{m_{up}} \psi_{k}(t-\tau_{m}) \, \Gamma_{mi}(t-\tau_{m}) \, \phi(m) dm  \\
\end{equation}
The quantity $\Gamma_{mi}$ is the matrix $Q_{mi}$ previously defined, with the diagonal elements (material already present in the star) 
set to zero.  
The conversion between the cosmic time $t$ and redshift $z$ is performed on the basis of the 
adopted cosmological model, which in our case is characterized by  
$\Omega_{m}=0.3$, $\Omega_{\Lambda}=0.7$  and $H_{0}=65 Km \, s^{-1} \, Mpc^{-1}$. \\
At a given redshift $z$, the total comoving mass density in the form of a given element (labeled $i$) 
produced in the galaxies of  $k-$th morphological type   
$\rho_{i,Tot}^{k} (M_{\odot} Mpc^{-3})$ is given by the integral: 

\begin{equation} 
\rho_{i,Tot}^{k}  (z) = \int_{0}^{z} \dot{\rho}_{i}^{k}(z') \, \frac{dt}{dz'} \, dz'  \\
\end{equation}
where $\dot{\rho}_{i}^{k}(z')$  is the production rate for the $i$-th element as a function of redshift for the galaxies of  $k-$th morphological type. 

As in CM04, 
the metals produced by spheroids and dwarf irregulars end up into three phases: some are locked up in stars plus remnants, some are contained in 
the ISM gas and the rest is expelled into  the IGM.  
At each instant, we can predict the comoving density of metals contained in each phase. 
To calculate the comoving mass density for the element $i$ and for  $k-$th galactic type  in stars, ISM gas and IGM, 
first we need to calculate the average stellar and ISM abundances predicted by our models for a typical spheroidal and dwarf irregular galaxy. 
These quantities are direct outputs of our chemical evolution models. 

Let us define $X_{i,*}^{k}(z)$ and $X_{i,g}^{k}(z)$ as the average 
abundances by mass locked in stars and remnants and in ISM gas, respectively, as a function of the redshift $z$ for the element $i$ in $k-$th galactic type.  
The stellar comoving density of the element $i$ in $k-$th galactic type is given by:\\
\begin{equation}
\rho_{i,*}^{k}(z)=X_{i,*}^{k}(z) \cdot \rho_{*}^{k}(z) 
\end{equation}
whereas the comoving density of the element $i$ in the ISM gas is given by:\\
\begin{equation}
\rho_{i,ISM}^{k}(z)=X_{i,g}^{k}(z) \cdot \rho_{g}^{k}(z) 
\end{equation}
$\rho_{*}^{k}(z)$ and $\rho_{g}^{k}(z)$ are the mass densities of stars plus remnants and ISM in the galaxies of the $k-$th morphological type, respectively, 
and are  calculated as: 
\begin{equation} 
\rho_{*}^{k}(z)=\rho_{B}^{k}(z) \cdot (M_{*}/L_{B})^{k}(z) \\
\end{equation}
and 
\begin{equation} 
\rho_{g}^{k}(z)=\rho_{B}^{k}(z) \cdot (M_{g}/L_{B})^{k}(z),\\ 
\end{equation}
where $(M_{*}/L_{B})^{k}(z)$ and 
$(M_{g}/L_{B})^{k}(z)$ represent the predicted stellar and gas mass to light ratios for  $k-$th galactic type, respectively.
  
The comoving density of a given element $i$ expelled by the galaxies of the $k-$th morphological type into the IGM is given by:\\
\begin{equation} 
\rho_{i,IGM}^{k}(z)=\rho_{B}^{k}(z) \cdot (M_{g}/L_{B})^{k}(z) \cdot \int_{0}^{z} (\dot{G_{i}})_{out}^{k}(z') \, \frac{dt}{dz'} \, dz'
\end{equation} 

The total mass density in the IGM for the element $i$ is then given by:
\begin{equation} 
\rho_{i,IGM}(z)= \sum_{k} \rho_{i,IGM}^{k}(z)
\end{equation} 
We assume that, immediately after the enrichment, the metals are distributed in the IGM homogeneously, i.e. 
we assume that  the IGM is ``well mixed''. This is probably a rough approximation, since the physical scale is very large, 
of $1 Mpc$ or more. 
Furthermore, not only metal enrichment, but also gas cooling and metal diffusion depend on the environment, i.e., local densities. 
These effects are not taken into account in our model, but we compensate by calculating the chemical enrichment in great detail. \\
To determine $\rho_{b,IGM}$ we have proceeded as in CM04. 
The first thing we need is a value for the total baryonic density, 
\begin{equation}
\Omega_{b} = \rho_{b}/\rho_{c}\\
\end{equation}
where $\rho_{c}=1.17 \cdot 10 ^{11} M_{\odot}/Mpc^{3}$ is the
critical density of the universe, calculated  for the value of
$h=0.65$ adopted here.  
For the baryonic density, here we assume the fiducial 
value of $\Omega_{b}=0.02 h^{-2}$ (Spergel et 
al. 2003), corresponding to  $\rho_{B}=5.55
\cdot 10^{9}  M_{\odot}/Mpc^{3}$.  To calculate the IGM baryon
density, we then subtract to this value the baryon density in galaxies 
(stars and ISM)  as calculated in CM04. 
The baryonic density in galaxies can be constrained by means of the observed luminosity density, as shown by Calura, 
Matteucci \& Menci (2004). 
They have shown that the assumption that the bulk of ellipticals formed at high redshift allows us to 
reproduce a large set of observational data.\\
We determine the abundances in the IGM in the following way:\\
\begin{equation}
[X_{i}/H]_{IGM}(z)= log \left( \frac{\rho_{i,IGM}(z)}{ \rho_{H,IGM}(z)} \right) -log(X_{i}/H)_{\odot}\\
\end{equation}
where $\rho_{i,IGM}(z)$ is the mass density of the element $i$ which has been ejected in the IGM by spheroids and dwarf irregulars 
and $\rho_{H,IGM}(z)$ is the comoving density of H in the IGM. 
To determine $\rho_{H,IGM}(z)$, for the IGM we assume a 
primordial composition with $^{4}He$ abundance of 0.25
(see Steigman 2005) and H abundance of 0.75.\\
$(X_{i}/H)_{\odot}$ is the solar abundance of the element $i$ relative to H\footnote{
The solar
abundances used in this paper are taken by  Anders \& Grevesse
(1989). } .\\ 
In the present work we do not consider metal production by pop III stars, 
and we assume that  only normal stars in high-redshift galaxies are 
the sources of the metals residing in the IGM. This assumption is motivated by the results from Matteucci \& Calura (2005), 
who have shown that pop III stars enrichment cannot account for the abundance ratios observed in the high redshift IGM. 
Furthermore, the critical metallicity necessary to enter a regime of normal star formation is  
$Z_{cr}=(10^{-6} - 10^{-4}) Z_{\odot}$ (Ferrara \& Salvaterra 2004, Yoshida, Bromm, Hernquist 2004),  
i.e. well below the values observed in the IGM considered here, namely $Z\ge 10^{-3} Z_{\odot}$. 


\begin{figure*}
\centering
\vspace{0.001cm}
\epsfig{file=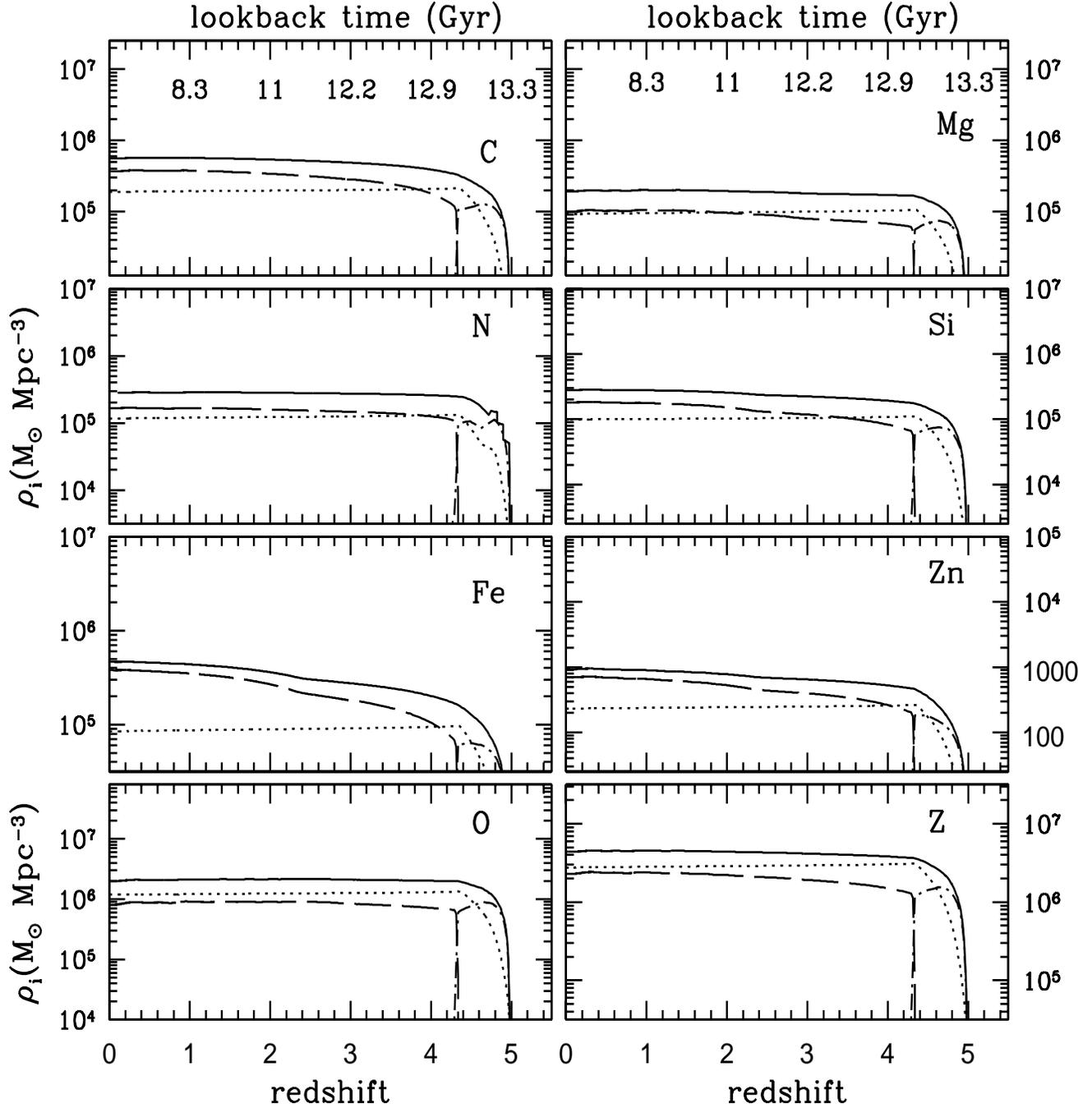,height=20cm,width=18cm}
\caption{
Predicted redshift (lower x-axis) and lookback-time (upper x-axis) 
evolution of the metals produced by spheroids (various elements). 
Solid lines: total mass density $\rho_{i}$ produced in the form of a given element $i$ 
(here the letter $Z$ indicates the total metallicity). Dotted lines: metal mass density locked up in stars. 
Dot-dashed lines: metal mass density contained in the ISM. Dashed lines: metal mass density ejected into the IGM. 
In this case, we assume that in all spheroids star formation starts at redshift $z_{f}=5$.
}
\label{metalsS}
\end{figure*}

\begin{figure*}
\centering
\vspace{0.001cm}
\epsfig{file=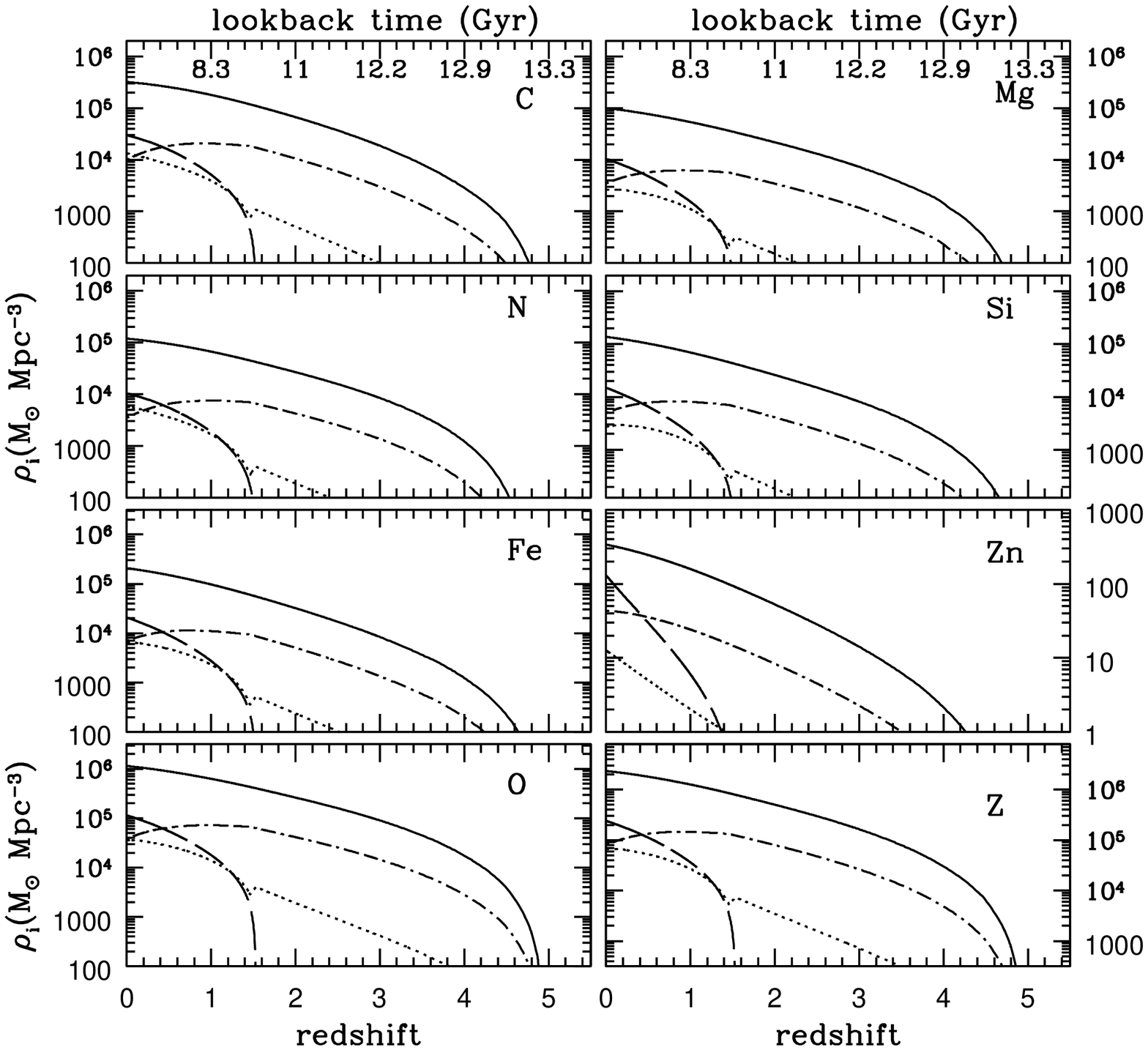,height=20cm,width=18cm}
\caption{
Predicted redshift (lower x-axis) and lookback-time (upper x-axis) 
evolution of the metals produced by dwarf irregulars (various elements). 
Solid lines: total mass density $\rho_{i}$ produced in the form of a given element $i$ 
(here the letter $Z$ indicates the total metallicity). Dotted lines: metal mass density locked up in stars. 
Dot-dashed lines: metal mass density contained in the ISM. Dashed lines: metal mass density ejected into the IGM. 
In this case, we assume that in all dwarf irregulars star formation starts at redshift $z_{f}=5$.
}
\label{metalsI}
\end{figure*}

\begin{figure*}
\centering
\vspace{0.001cm}
\epsfig{file=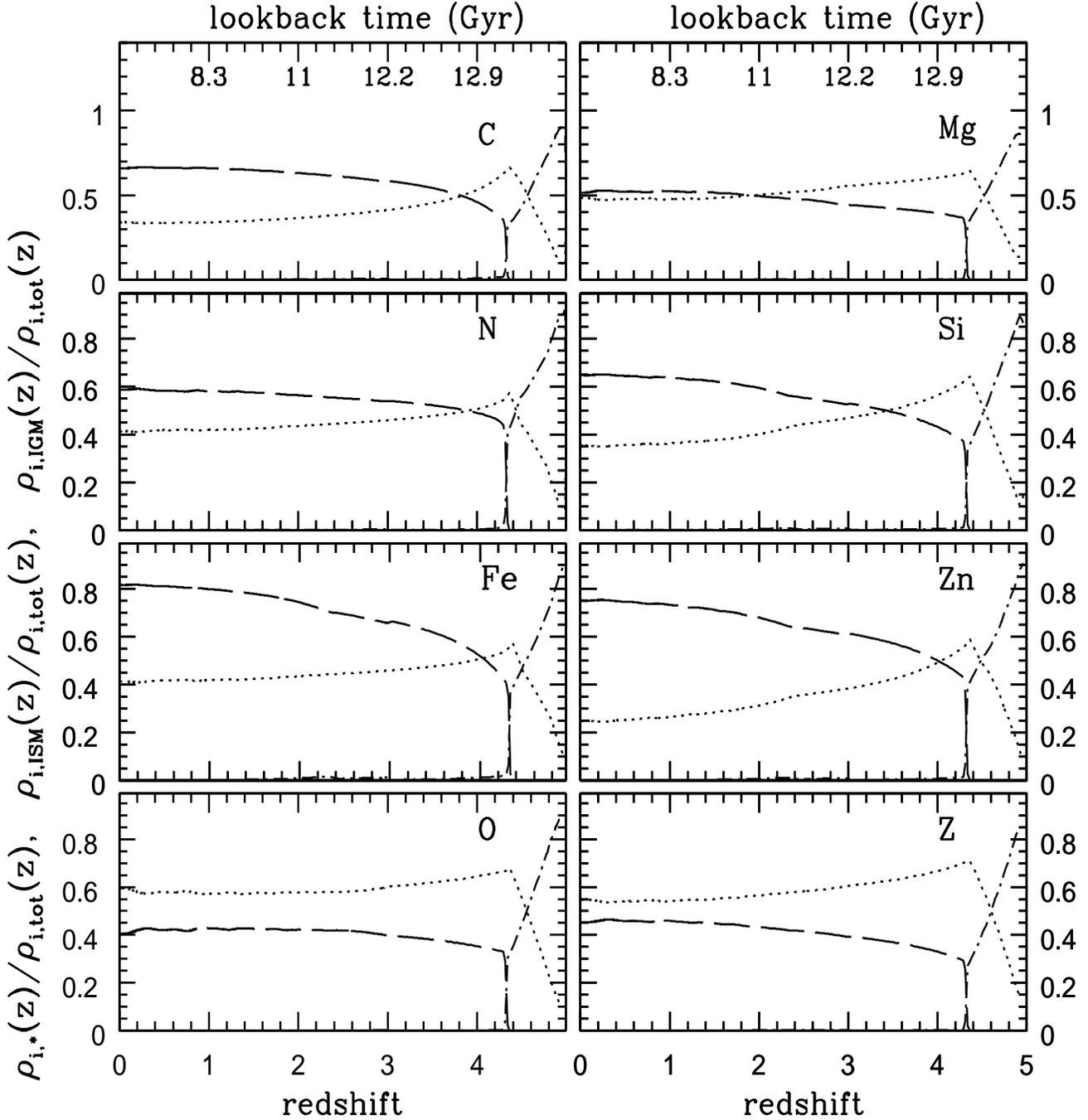,height=20cm,width=18cm}
\caption{
Predicted redshift (lower x-axis) and lookback-time (upper x-axis)  
evolution of the metal fractions, normalized to the total mass in the form of a given element,  
locked up in galactic spheroids in stars (dotted lines), 
in the ISM (dot-dashed lines), in the IGM (dashed lines) for various elements.   
In this case, we assume that in all spheroids star formation starts at redshift $z_{f}=5$.
}
\label{fractionsS}
\end{figure*}

\begin{figure*}
\centering
\vspace{0.001cm}
\epsfig{file=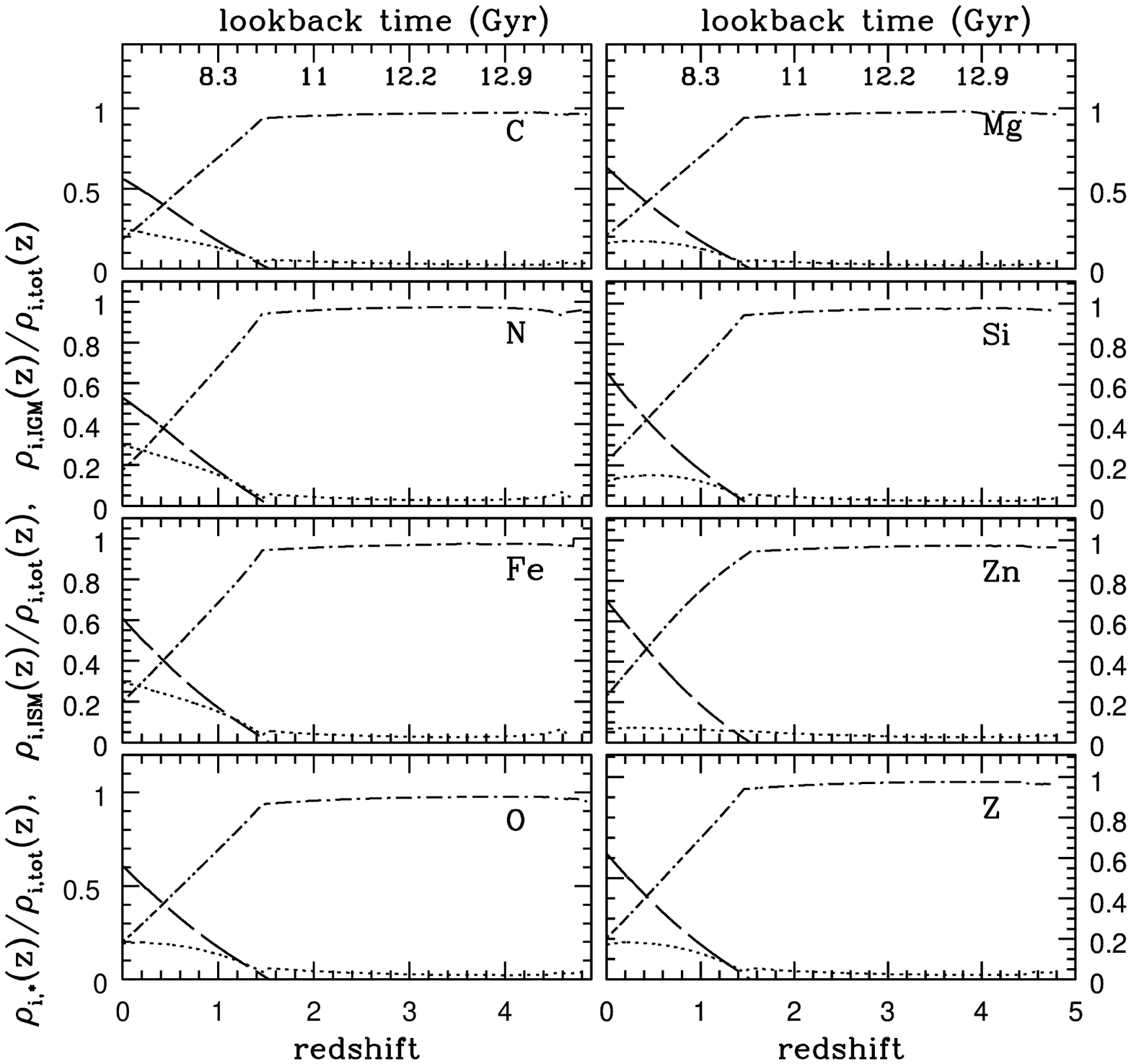,height=20cm,width=18cm}
\caption{
Predicted redshift (lower x-axis) and lookback-time (upper x-axis)  
evolution of the metal fractions, normalized to the total mass in the form of a given element,  
locked up in dwarf irregular galaxies in stars (dotted lines), 
in the ISM (dot-dashed lines), in the IGM (dashed lines) for various elements.   
In this case, we assume that in all dwarf irregulars star formation starts at redshift $z_{f}=5$.
}
\label{fractionsI}
\end{figure*}

\section{Results}

\subsection{The evolution of the cosmic metal share}  
In Figures ~\ref{metalsS} and ~\ref{metalsI} we show the predicted metal comoving densities 
contained in the stars, in the  ISM 
and those ejected into the IGM  as functions of redshift, as produced by spheroids and dwarf irregulars, respectively. 
In Figures ~\ref{fractionsS}  and ~\ref{fractionsI}  
we show the evolution of the metal fractions in the ISM, IGM and in the stars,
i.e., at each redshift,  the ratio between the metal mass density in each phase (stars, IGM and ISM) and the total one for 
any given element, as produced by spheroids and dwarf irregulars, respectively. 
Here, we show only the results obtained by assuming that both spheroids and dwarf irregulars started forming stars at redshift $z_{f}=5$. 
Figure~\ref{metalsS} and ~\ref{metalsI} provide information on the values of the total comoving mass densities, 
expressed in $M_{\odot} Mpc^{-3}$, obtained by summing up the various phases 
 at any redshift. On the other hand, 
%
%
Figures~\ref{fractionsS} and ~\ref{fractionsI}  are useful to see how the fractions of metals contained in the various phases 
change with the redshift. \\ 
In the case of galactic spheroids (Figures~\ref{metalsS} and ~\ref{fractionsS}), we see that 
elements produced on different timescales show different behaviours. 
Elements produced by low and intermediate mass stars, such as C and N, are  mainly produced on long timescales ($> 1 Gyr$), with their bulk being produced 
after the end of star formation. \\
Elements produced by massive stars  ($M>8 M_{\odot}$) on short timescales ($\le 0.03 $Gyr), such as O and Mg, are produced during the star 
formation epoch. At the present time, nearly half of these elements is locked up into the stars.\\
Fe is an elements produced by both type Ia and II SN explosions, but in different proportions: 
3/4 of the Fe originates in type Ia SNe, whereas the remaining 1/4 comes from type II SNe 
(Matteucci \& Greggio  1986). 
According to the nucleosynthesis prescriptions adopted here, 
Zn is produced in a substantial way in SNe Ia. 
Therefore, both Fe and Zn  are then massively produced after the starburst.\\
Si is an element produced both by type II and type Ia SNe and in similar proportions. 
At the present time, nearly 30 $\%$ of all the Si produced in spheroids is locked up into stars.\\
In the case of dwarf irregular galaxies, the total metal density and stellar metal density 
(solid and dotted lines in Figure~\ref{metalsI}, respectively) 
is a slowly increasing function throughout all their history. This is due to a continuous and low star formation rate characterizing 
these objects. The ISM metal densities (dot-dashed lines  in  Figure~\ref{metalsI}) increase until the time at which  
the galactic wind develops, i.e. for the model used here, 3.6 Gyr after the beginning of star formation, 
corresponding to redshift $z=1.5$. At this redshift, the metal densities in the ISM start to decrease, whereas  
the metal densities in the IGM start to increase.

\subsection{Chemical abundances in the IGM} 
\begin{figure*}
\centering
\vspace{0.001cm}
\epsfig{file=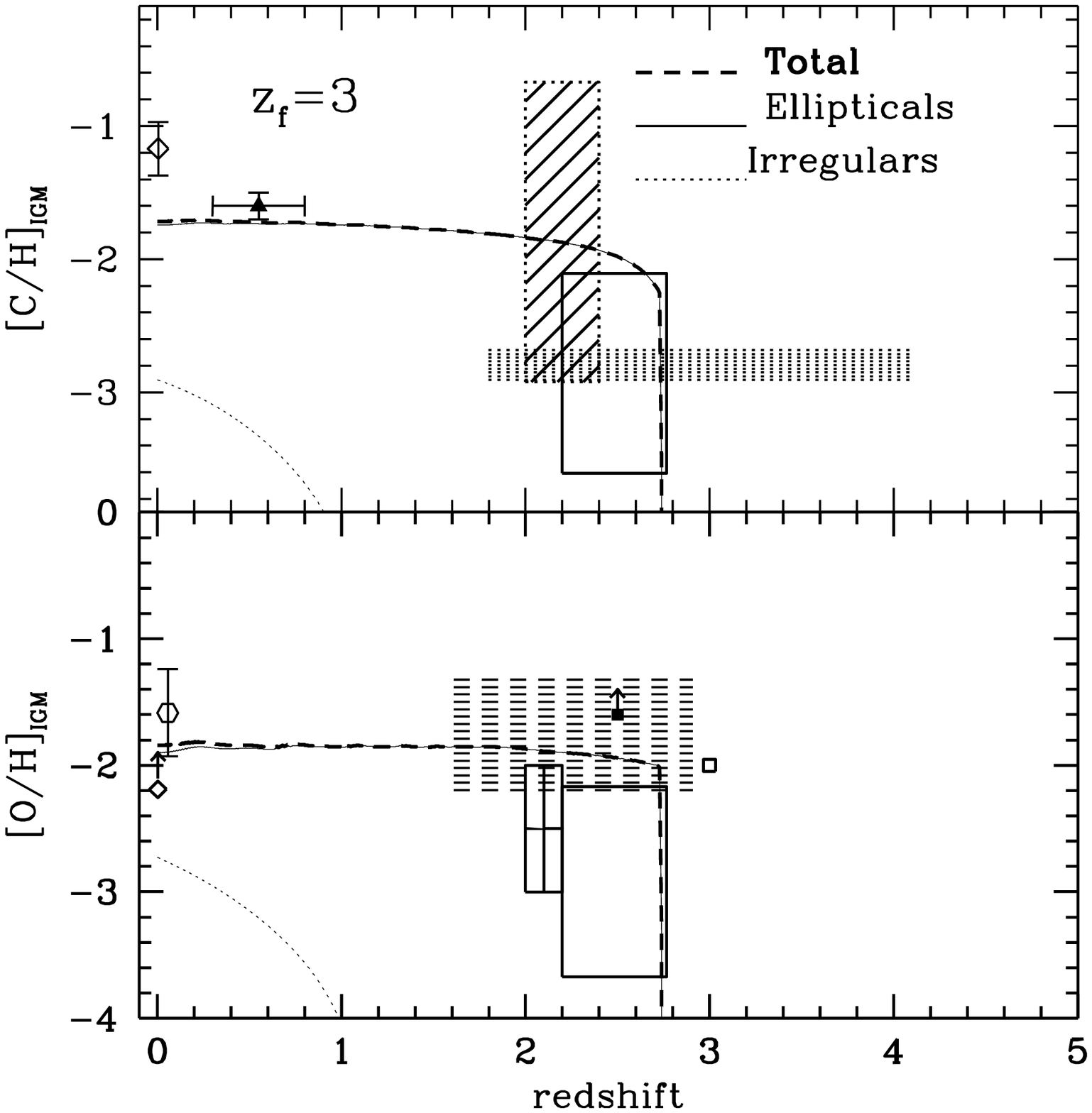,height=10cm,width=10cm}
\caption{
Evolution of the C (upper panel) and O (lower panel) abundances in the IGM as measured by various authors and as predicted by our models 
by assuming that the star formation in spheroids and dwarf irregulars starts at $z_{f}=3$. 
\emph{Thin solid lines}: contribution by spheroids. \emph{Thin dotted lines}: contributions by dwarf irregulars. 
\emph{Thick dashed lines}: total abundances. 
In both panels, the observational data indicated with squared and shaded regions have been derived by taking into account different ionizing UV 
backgroud shapes. 
\emph{Data in upper panel}: open diamond: 
Tripp et al. (2002). Solid triangle: Barlow \& Tytler (1998). 
Light-hatched area: Bergeron et al. (2002). Square box: Simcoe et al. (2004a). Dotted area: 
Schaye et al. (2003). 
\emph{Data in lower panel}: open diamond: 
Tripp et al. (2002). Open hexagon: Shull et al. (2003). Solid square: Simcoe et al. 2002.
Empty square: Dav\'e et al. (1998). 
Dashed area: Telfer et al. (2002). Crossed box: Carswell et al. (2002).  Square box: Simcoe et al. (2004a).
}
\label{abundaz3}
\end{figure*}

\begin{figure*}
\centering
\vspace{0.001cm}
\epsfig{file=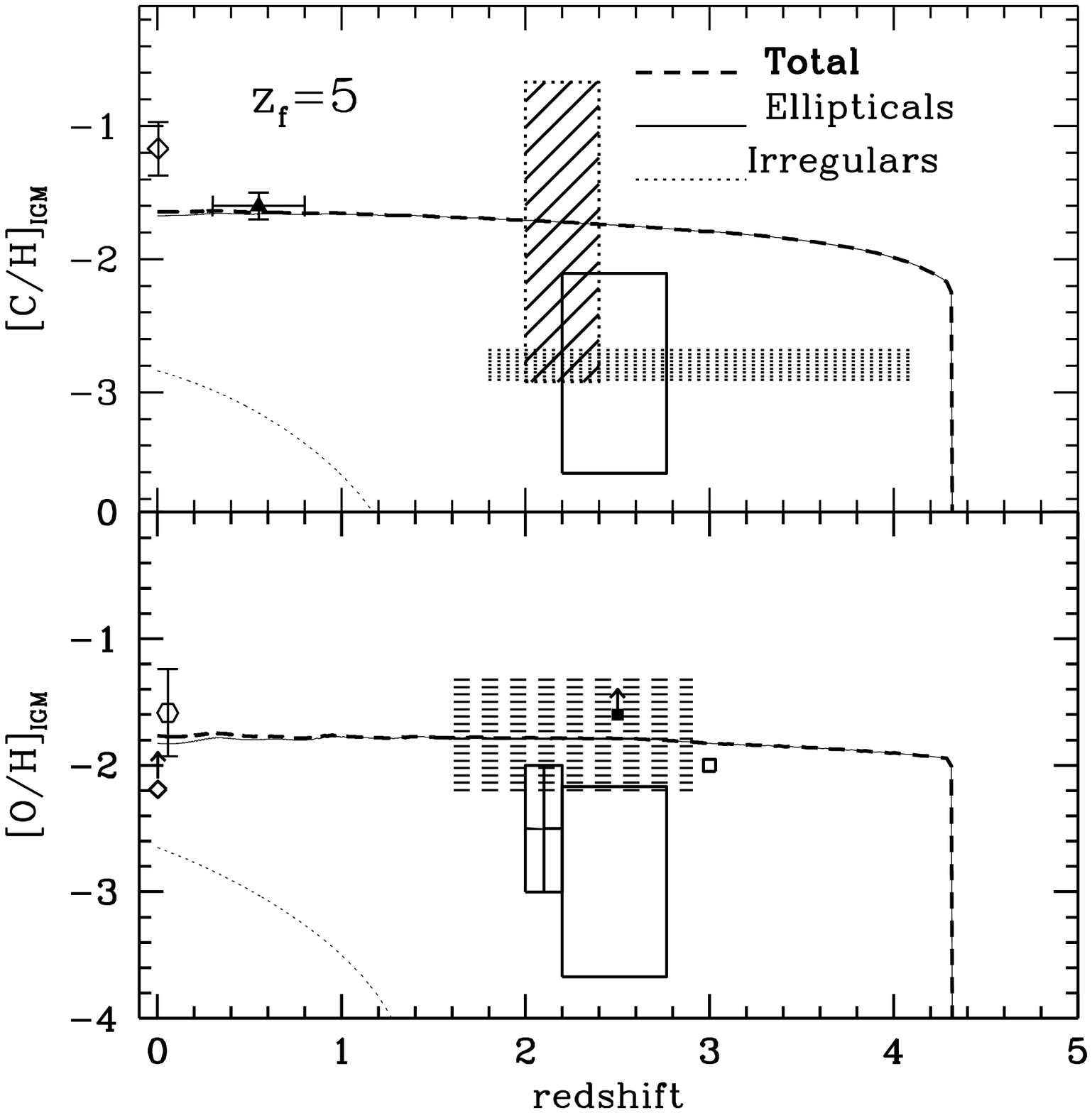,height=10cm,width=10cm}
\caption{Evolution of the C (upper panel) and O (lower panel) abundances in the IGM as measured by various authors and as predicted by our models 
by assuming that the star formation in spheroids and dwarf irregulars starts at $z_{f}=5$. 
Lines and data as in Figure~\ref{abundaz3}. 
}
\label{abundaz5}
\end{figure*}

The abundances in the IGM are determined by means of the metal absorption lines observed in the quasar 
(QSO) spectra. 
In these spectra, the presence of the IGM is indicated by the multitude of individual 
Lyman-$\alpha$ (Ly$\alpha$) absorption 
lines bluewards of the Ly$\alpha$ emission peak of the QSO. 
These absorption lines originate in ionized gas clouds 
with typical column densities in the range $10^{16} \ge N(HI) \ge 10^{12} cm^{-2}$  (Pettini 2004, Bechtold 2003), 
located in the lines of sight of the QSOs. 
The abundances in the Ly$\alpha$ forest are usually determined for C, O and Si. 
Many are the uncertainties in the determination of the abundances and concern the modeling 
of the IGM conditions 
(Simcoe et al. 2004b, Carswell 2004). Major uncertainties are also the nature of the UV photon flux, 
i.e. whether visible or dust-obscured QSOs 
(Rees \& Setti 1970, Pei \& Fall 1995) or star forming galaxies (Haardt \& Madau 1996, 
Kim, Cristiani, D'Odorico 2002, Bianchi, Cristiani \& Kim 2001) 
represent the principal sources of UV photons. 
Observationally, the of the Oxygen abundance $[O/H]$ is derived by using measurements of $O_{VI}$ and $H_{I}$ absorptions, according to:\\
\begin{equation} 
[O/H]_{obs}=log(\frac{N_{O_{VI}}}{N_{H_{I}}})+log(\frac{f_{H_{I}}}{f_{O_{VI}}})-log(\frac{O}{H})_{\odot}\\
\end{equation}  
where $N_{O_{VI}}$ and $N_{H_{I}}$ are the column densities of the chemical species $O_{VI}$ and $H_{I}$, respectively. 
$f_{O_{VI}}$ and $f_{H_{I}}$ are the ionization fractions, i.e. the ratios between the column density 
of the ionized species  
$O_{VI}$ and $H_{I}$ with respect to the neutral ones (Simcoe et al. 2004a). The same method may be applied 
to calculate [C/H] and [Si/H] from the $C_{IV}$ and $Si_{IV}$ absorption lines, respectively. 
The main challenge in determining the abundances is an accurate estimate of the ionization fractions, 
depending on all 
the physical properties of the IGM  discussed above.\\ 
In 
figures~\ref{abundaz3} and ~\ref{abundaz5} we show the evolution of the 
intergalactic C (upper panel) and O (lower panel) 
abundances 
as predicted by our models and 
as observed by various authors, having assumed that all galaxies started to form 
stars at $z_{f}=3$ and $z_{f}=5$, respectively. 
For this comparison we consider only 
these two elements  as  they represent the ones with the largest number 
of measures. 
 Some of the observations considered here have been derived by testing different shapes 
of the ionizing UV backgrounds (dominated by QSO or a mixture between galaxies and QSOs). 
These data are represented in Figures~\ref{abundaz3} and~\ref{abundaz5} by squared and shaded regions. 
Data derived by assuming one single type of UV background are represented by points with error bars. 

In these figures we show both the contributions by spheroids (thin solid lines) and dwarf irregulars (thin dotted lines) to 
the total IGM abundances, along with the sum of the two (thick dashed lines).  
For both elements, it is worth to note that the contribution by dwarf 
irregular galaxies to the IGM appears negligible with respect to spheroids. 
Dwarf irregulars contribute to the $6 \%$ of the C mass density and to the $12 \%$ 
of the O mass density present in the IGM at the prsent time. 
This is an important result, which is in agreement with the ones already found 
in Calura \& Matteucci (2003), where we showed that these   
objects have played a negligible role in the cosmic star formation history.
In the case of [C/H], in both figures~\ref{abundaz3} and ~\ref{abundaz5} 
we note that at high redshift the predicted total IGM abundances are 
in agreement with the value by 
Bergeron et al. (2002, light-hatched area), 
but overestimate the IGM [C/H] 
abundance  as measured by  Simcoe et al. (2004a, square box) and Schaye et 
al. (2003, dotted area). 
If we take the observational data at face value and neglect all their uncertainties, we note that our predictions calculated 
by assuming $z_{f}=3$ 
seem in better agreement with the measures than the ones calculated by assuming $z_{f}=5$. 
The predictions  are in very good agreement with the estimates by Barlow \& Tytler (1998, shaded area) at redshift $z \sim0.5$, 
but underestimate the value by Tripp et 
al. (2002, empty diamond) at redshift zero by $\sim 0.4$ dex.  The one by Tripp et al. (2002) is the only local observational estimate of the C abundance in the 
IGM, and it has been derived by considering only a UV background dominated by QSOs (Haardt \& Madau 96). 
A softer UV background, including also galaxies, would produce significantly lower C abundances (see Schaye et al. 2003). \\
The most reliable IGM metallicity tracer is represented by O since, at variance with C, 
it is observable down to the lowest density regions of the universe, characterized by HI column density values of 
$N_{HI} \le 10^{14.5} cm^{-2}$. 
Unfortunately, also the determination of the abundances of O suffers the same uncertainties as C. 
Furthermore, O is detected by means of the $O_{VI}$ absorptions, which arise in the Lyman $\alpha$ forest. 
For this reason, the measure of its column density is contaminated also by blending effects.\\ 
The [O/H] in the IGM (figures~\ref{abundaz3} and ~\ref{abundaz5}, lower panels) is roughly accounted for by our 
predictions both at high and low redshift.  
We note that the 
observations show no strong evolution in the O abundances between 
$z=2.5$ and $z=0$, suggesting that the IGM was substantially  enriched with 
metals at very high redshift.  
At high redshift ($z>3$), our predictions are consistent with the value 
by Dav\'e et al. (1998, empty square), which has been provided with no error bar. 
However, if the uncertainties are the same as for the data by the other 
authors, i.e. $\delta \ge 0.5$ dex, we can say that this measure is more consistent with our predictions 
calculated by assuming a redshift of formation $z_{f}=5$. 
Our predictions overestimate the values observed by Carswell et al. (2002 crossed regions in 
figures~\ref{abundaz3} and ~\ref{abundaz5}) and Simcoe et al. (2004a, large square box), but are 
in very good agreement with the data by Telfer et al. (2002, dashed region). 
The only measured abundance being slightly higher than our predictions is the value by  Simcoe et
al. (2002) at $z\sim 2.5$. Although being underestimated by our models and higher than any other value,  
this measure represents a lower limit and it is unlikely to represent the average IGM O abundance. 
It is likely that this object represents a galactic wind in action, or 
a proto-cluster region: in fact, the intergalactic
systems associated to those absorptions are  located in highly overdense
environments, where the gas density is particularly large and the
star formation and metal  production rates could be enhanced.\\
Finally, the data at redshift zero by Shull et al. (2003, open hexagon) and the lower limit by Tripp et al. (2002) are 
in very good agreement with our predictions. \\
If the main responsible of the IGM enrichment are the big spheroids, this
lack of evolution in the IGM O abundance is another independent
indication that the largest spheroids have formed at very high
redshift,  consistently with many other observational evidences such
as the fundamental plane (Renzini \& Ciotti 1993), the Mg-$\sigma_{*}$ relation (see Bernardi et al. 2003 and references therein) 
and the color-magnitude relation (Bower, Lucey \& Ellis 1992). \\ 
 From the analysis described in this section, we conclude that our predictions can account for the observed 
evolution of the [O/H] in the IGM, being consistent with most of the existent data. 
In the case of [C/H] in the IGM, our predictions overestimate 2 of the 3 measures quoted above. 
This could be due to inadequate ionization 
corrections applied to the   $C_{IV}$ column density estimates.  
It is well known that the C abundances are based primarily on   $C_{IV}$ observations, which probe the densest 10\% of the baryons in the universe and 
the 1\% of the cosmic volume (Simcoe et al. 2004b). In regions where the density is near the universal mean,   $C_{IV}$ loses its sensitivity as a metallicity tracer, 
since most of the C has been ionized to $C_{V}$ and higher levels. 
In those regions, we could be still missing the bulk of the C. 
If this is true, all the observational data should be regarded as lower limits. On the other hand, although more 
difficult to detect owing to blending effects, in these very low density regions $O_{VI}$ does not lose its reliability as a metallicity tracer, owing to 
its ionization potential and the high O abundance (Carswell et al. 2002).\\ 
Another possible reasons for the discrepancies between predictions and data for both C and O could be the 
fact that the IGM is not well mixed as assumed in the present work. 
The data could be biased and could not represent the average values of the whole universe. 
Finally, a high fraction of the metals could not b observable, residing  at higher temperature in the intra-cluster hot gas, 
or in warm gas. \\ 

\subsection{The IMF and the abundance ratios in the IGM} 
\begin{figure*}
\centering
\vspace{0.001cm}
\epsfig{file=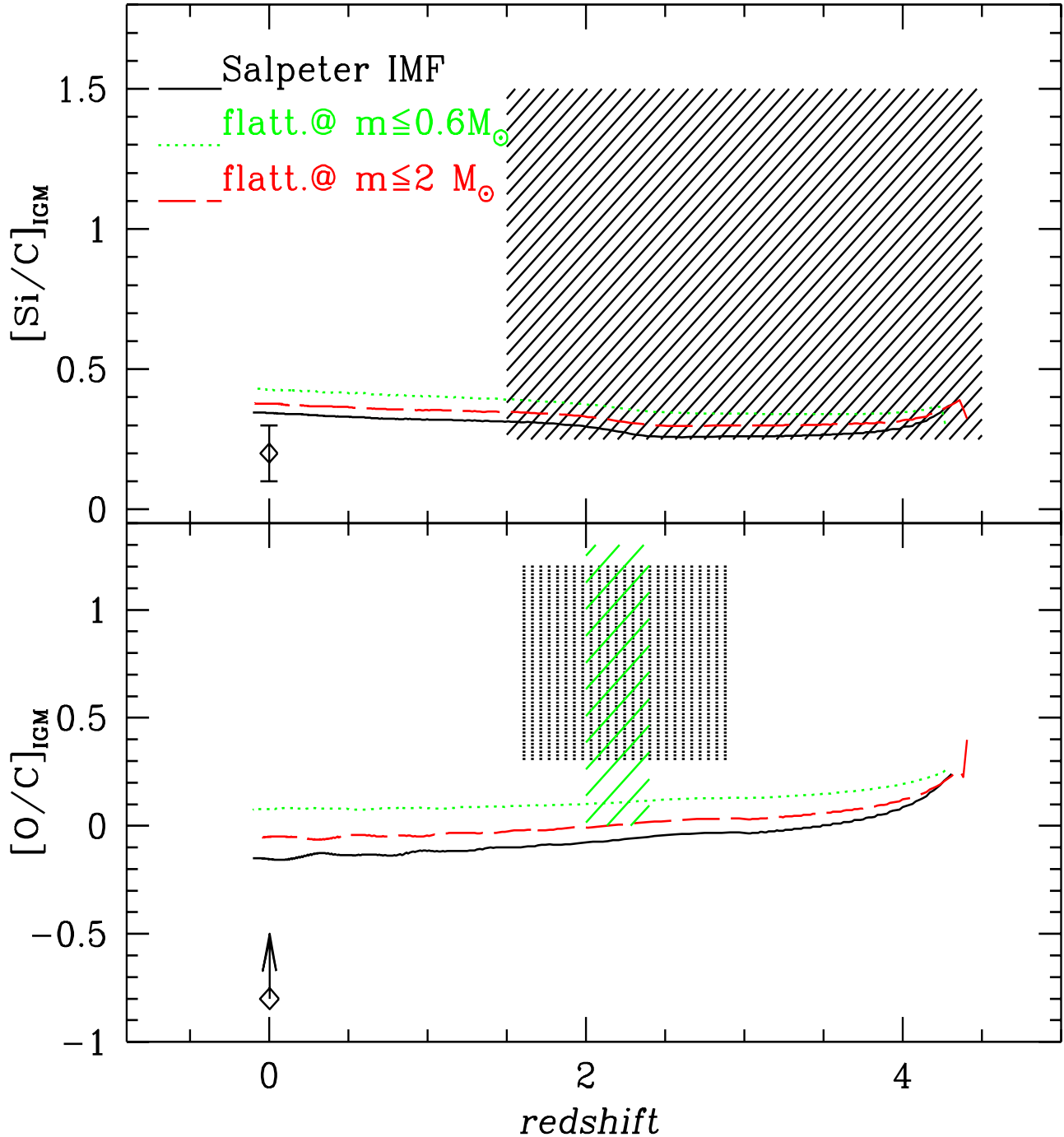,height=20cm,width=18cm}
\caption{Evolution of the [Si/C] and [O/C] 
abundance ratios in the IGM as predicted by our models for different IMFs and as measured by various authors. 
\emph{Upper panel}: solid line: Salpeter IMF. Dotted line: IMF flatter ($x=0.35$) than the Salpeter at masses $m\le 0.6 M_{\odot}$.  
Dashed line: IMF flatter ($x=0.35$) than the Salpeter at masses $m\le 2  M_{\odot}$. 
Observations: shaded area: Aguirre et al. (2004); open diamond: Tripp et al. (2002). 
\emph{Lower Panel}: curves as in upper panel. 
Observations: shaded area: Telfer et al. (2002). Light-hatched area: Bergeron et al. (2002). Open diamond: Tripp et al. (2002). 
In this case, we assume that star formation in spheroids starts at $z_{f}=5$. 
}
\label{abundarat5}
\end{figure*}

The possibility of a non-standard Salpeter IMF in 
various environments has already been suggested by various authors. 
In the disc of the Milky Way, the stellar IMF is steeper 
than the one typical of young open clusters (Kroupa 2004; Scalo 2004). 
This is supported by results from chemical evolution models, 
finding that in general a Salpeter-like IMF tends to overestimate the 
metal abundances  observed in the local disc (Romano et al. 2004; Portinari et al. 2004). 
In elliptical galaxies, it has been noted that the standard Salpeter IMF tends to 
overestimate the average $M/L_{B}$ and that the best results are obtained 
by flattening the IMF at masses $\le 1 M_{\odot}$ (Renzini 2004).\\
From a theoretical point of view, 
the main parameters affecting the abundance ratios are the adopted stellar yields and 
the stellar IMF. 
Here, we assume a fixed set of yields 
and we try to modify the IMF, in order to investigate also the effects of a 
non-standard  Salpeter-like IMF on the predicted abundance ratios. 
In figure~\ref{abundarat5}, we compare 
observational estimates of the [Si/C] (upper panel) and [O/C] (lower panel) with our predictions, by assuming that 
star formation in spheroids started at redshift $z_{f}=5$.  
We are showing predictions calculated assuming various types of IMF:  
a standard Salpeter (solid lines) and a modified Salpeter   
with flatter slope, $x=0.35$ instead of $x=1.35$,  below two mass-cuts,
$0.6 M_{\odot}$ (dotted lines) and $2.0 M_{\odot}$ (dashed lines), respectively. \\
Recently, Aguirre et al. (2004) have carried out a detailed study of 
silicon,  in the IGM in the redshift range  $z 
\sim 2-4$.   By using the results presented by Schaye et al. (2003),
who measure carbon abundance in the same redshift range, and assuming
different  UV background types,  they have provided an observational
estimate of the [Si/C] ratio in the IGM.  The [Si/C] observed at high
redshift by Aguirre et al.  (2004, shaded area in
Figure~\ref{abundarat5}, upper panel) spans a broad range,  from
[Si/C]=0.25, obtained with a UV background (UVB) including galaxies,
to [Si/C]=1.5 in the case of UVB only due to QSOs.  As a redshift
range, we have assumed the one of the QSOs studied by Aguirre et
al. (2004), i.e. $1.5 \le z \le 4.5$. 
A local estimate of the [Si/C] ratio 
in the IGM is provided by Tripp et al. (2002),  who studied the chemical
abundances in two Ly$\alpha$ clouds in the vicinity of the Virgo 
supercluster, finding $[Si/C] \sim 0.2 \pm 1$ (empty diamond in upper panel of figure~\ref{abundarat5}). \\
Our predictions are consistent with the range of values provided by Aguirre et al. (2004). 
We note that no choice of the IMF allows us to have values for the [Si/C] ratio at high redshift 
larger than $[Si/C]\sim 0.4$ dex.\\
At redshift zero, the predicted  [Si/C] values for various IMFs are higher than the value observed by 
Tripp et al.  (2002). More observations with a detailed study of the effect of different ionizing 
UV background shapes would be very useful in order to understand the nature of this discrepancy. \\  
In Figure~\ref{abundarat5}, lower panel, we show the
evolution of the [O/C] ratio in the IGM, as predicted by our models
and as  observed by various authors.  The data for the [O/C] ratio in
the IGM at high redshift ($z > 1.5$) are from Bergeron et al. (2002, 
Light-hatched area)
and Telfer et al. (2002, shaded area). At redshift zero, 
we use the values by Tripp et al. (2002, empty diamond),
where [O/H] represents a lower limit.  
At high redshift, our predictions lie outside of the range indicated by Telfer et al. (2002), but are marginally consistent with the data by Bergeron 
et al. (2002). 
In general, the flatter is the IMF in the low mass regime, the higher is the obtained [O/C] value and the better is the agreement 
between observations and predictions. 
 Globally, our results indicate that it is hard to reproduce the abundance ratios [O/C] and [Si/C] 
observed in the IGM at redshifts larger than 2. 
 One way to match the observations would be to adopt a top-heavy IMF in spheroids, 
but in CM04 we have already shown that such a choice would lead to a severe overestimation of the local metal budget. 
On the other hand, the adopted stellar yields could be inappropriate.  
 The predicted carbon abundance is too large. In order to reproduce the  observed [Si/C] and [O/C], it would be necessary 
to decrease ad-hoc the C yields adopted in this work. However, it is worth to stress that for C, 
the nucleosynthesis 
yields adopted here allow us to reproduce the abundance pattern observed in the Galactic stars with good accuracy (Chiappini et al. 2003). 
With such an ad-hoc  modification of the C yields, it could be hard to reproduce the local C abundance. \\
Another possible solution could be that a non-negligible fraction of C is locked up  into dust grains. This would cause an 
underestimation of the actual IGM  C abundance. The treatment of dust formation in the chemical evolution equations could be 
very useful to investigate this possibility (Calura et al., in preparation).   
Finally, inhomogeneities and local effects could play an important role, not taken into account in our models.

\subsection{Escape of metals in IGM and stall radius} 

\begin{figure*}
\centering
\vspace{0.001cm}
\epsfig{file=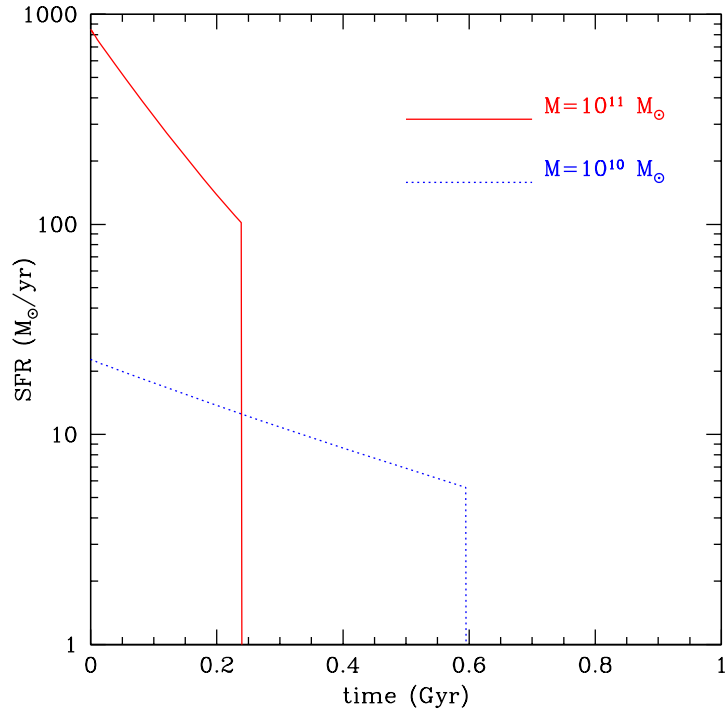,height=10cm,width=10cm}
\caption{Star formation rate, expressed in $M_{\odot}/yr$, for an elliptical galaxy of baryonic mass $M=10^{11} M_{\odot}$ (solid line) 
and $M=10^{10} M_{\odot}$ (dotted line), as a function of time. 
}
\label{SFR}
\end{figure*}

An important aspect of the IGM metal enrichment concerns how far 
the metals can travel in the space once they are ejected, and whether they can reach the lowest 
density regions of the IGM where metals are detected. 
Ferrara, Pettini \& Shchekinov (2000) have performed a theoretical study of the mixing of the metals in the early IGM. 
By means of a semi-analytical model for SN feedback, they concluded that SN-driven winds, or blowouts, fail to 
distribute the metals over volumes large enough to pollute the whole IGM.  
Both dwarf and giant galaxies eject metals over maximum distances $R_{max}$ of $\le 100 kpc$, a factor of 10 smaller 
than the average physical separation between massive objects at high redshifts, which is of the order of $1 Mpc$. 
This means that we should expect that regions outside the radius $R_{max}$ have a primordial chemical composition. 
Ferrara et al. (2000) conclude that SN-driven galactic winds are inadequate to explain the ubiquitous presence of metals in the IGM, 
and that additional mechanisms must be at work to account for the efficient mixing of the metals on scales of $\sim 1 Mpc$.\\
On the other hand, in a more recent paper, Tumlinson \& Fang (2005) have studied the distribution of the intergalactic $O_{VI}$ absorbers in the 
local IGM. 
They concluded that, in order to account for the observed distribution of metals in the local IGM, 
massive galaxies have to enrich regions of $\sim 0.5 \, - \, 1 Mpc$. \\ 
On the basis of simple physical arguments, we attempt to calculate the maximum distance, or ``stall  radius '' $R_{s}$ 
which could by reached by the ejected metals.\\
We assume that a spheroid of baryonic mass $M_{LBG}=10^{10} M_{\odot}$ and $M_{ULIRG}=10^{11} M_{\odot}$ 
represents a typical Lyman break galaxy (LBG) and ULIRG, respectively. 
According to the inverse wind scenario, we assume that the star formation efficiency increases with the galactic mass: 
for the $10^{10} M_{\odot}$ and $10^{11} M_{\odot}$ galaxies, we assume star formation efficiencies $\nu_{LBG}= 3 $ Gyr$^{-1}$ and 
$\nu_{ULIRG}= 11 $ Gyr$^{-1}$ (Matteucci 1994, Pipino \& Matteucci 2004). \\
In Figure~\ref{SFR}, we show the SFR, expressed in $M_{\odot}/yr$, for the models representing the ULIRG (solid line) and the LBG (dotted line), 
as a function  of time. 
According to our results, a  $10^{11} M_{\odot}$ elliptical develops a wind in $0.23$ Gyr, with average values for the SFR of 
$\sim 10^{11} M_{\odot} / 2.3 \cdot 10^{8} yr \simeq 400  M_{\odot}/yr $. 
The SFR of the most massive elliptical galaxies can reach very high peak values (up to $\sim 1000 M_{\odot} yr^{-1}$), similar 
to the ones observed in some 
SCUBA (Ivison et al. 2000) and luminous infrared galaxies (Rowan-Robinson et al. 1997, 2000). \\
On the other hand, a  $10^{10} M_{\odot}$ spheroid develops a wind in $\sim 0.6$ Gyr, 
hence the average star formation rate is $\sim 17 M_{\odot}/yr$.  \\
Here, the basic assumptions for the ISM heating by SNe and stellar winds 
are the same as described in various papers (Pipino et al. 2002, Pipino et al. 2005). 
As in Pipino et al. (2002), we assume that every SN releases an energy $E_{0}=10^{51} erg$, 
$\sim 20 \%$ of which goes into thermal heating of the ISM, after taking into account SN remnant cooling. 
We assume that all the thermal energy is carried out by the wind. In this case, at each timestep the energy lost in the galactic wind is 
$\Delta E = E_{th}$, where $E_{th}$ is given by equation (1). 
If the quantity $\Delta M$ represents the amount of matter ejected through the galactic wind at each timestep, 
we can then calculate the wind velocity according to $v_{w} = (2\Delta E /\Delta M)^{1/2}$ (Pipino et al. 2005). 
With the prescriptions adopted here, immediately after the starburst we obtain wind velocities of 
$v_{w,LBG} \sim 500 Km/s$ and $v_{w, ULIRG} \sim 1400 Km/s$ in the case of a typical LBG and ULIRG, respectively. 
Both values are compatible with the ones measured in real LBGs (Pettini et al. 2002, Lehmer et al. 2005), local starbursts (Martin 2005) and  
LIRGs/ULIRGs  (Veilleux 2003, Veilleux et al. 2005).\\
To give an estimate of the stall radius $R_{s}$, we assume that the wind propagates as a spherical bubble in a uniform medium.  
We require that the outflow ram pressure balances the IGM pressure $p_{IGM}$ (Ferrara et al. 2000), obtaining: 
\begin{equation} 
R_{s}=\left( \frac{\dot{M}_{w} v_{w}}{4 \pi p_{IGM}} \right) ^{1/2} 
\end{equation} 
where $\dot{M}_{w}$ is the mass loss rate of the galaxy. 
According to our predictions, 
immediately after the onset of the wind the mass loss rate is $\sim 1.9M_{\odot}/yr$ and $9 M_{\odot}/yr$ 
in the case of the LBG and ULIRG, respectively.  
The pressure of the IGM is given by: 
\begin{equation}
p_{IGM}=p_{*}\left[ (1+z)/200     \right]^{5}
\end{equation}
where $p_{*}/K_{B} = 1500 \, h^{2} cm^{-3} K$ is the IGM pressure at $z=200$ (Ferrara et al. 2000). 
A fundamental quantity to calculate $p_{IGM}$ is the IGM temperature $T_{IGM}$. 
The structure of the IGM is likely to be multiphase, with a colder medium at temperatures  $T= 10^{3.5 - 5} K$ 
(Danforth \& Shull 2005, Stocke, Shull \& Penton 2004). Here, we assume for $T_{IGM}$ the value in-between these two media, i.e. 
$T_{IGM}=10^{5} K$.  
With these hypothesis, at redshift $z=4$ for the stall radius 
we obtain a value for $R_{s, LBG}=0.77 Mpc$  and  $R_{s, ULIRG}= 2.8  Mpc$ for LBGs and ULIRGs, respectively. 
We conclude that, on the basis of these very simple arguments, 
the metals ejected by massive galaxies can reach very long distances, being in principle able to account for the observed ubiquity of the 
metals in the diffuse IGM. 
Furthermore, it is worth to note that, with such values for the wind velocity and assuming this quantity to be constant, the ejected matter would 
be able to move to the required 
distances of $\sim 1 Mpc$ on timescales of  $\sim 1 Mpc /v_{w, LBG}=1.95 Gyr$ and     $\sim 1 Mpc /v_{w, ULIRG}=0.7\, Gyr$, 
in the case they are ejected by a LBG or a ULIRG, respectively.
The most massive galaxies, experiencing the most intense starburst in the universe, could be responsible for the enrichment of 
the lowest density regions of IGM, which are also the most distant from galaxies. 
It has been found recently that the role of ULIRGs in the cosmic star formation increases rapidly at redshift $z>1.5$, and that 
at redshift $z\sim 3$ they completely dominate the SFR density (P\'erez-Gonz\'alez et al. 2005). 
Furthermore, the presence of very massive starbursts is detected up to $z\sim 5.5$ (Staguhn et al. 2005),  
along with the presence of very metal-rich massive quasars  (Dietrich et al. 2003, Iwamuro et al. 2004).  
On the light of these evidences, it is likely that these objects dominate also the metal enrichment of the IGM at redshift $z\sim 3-5$. 
In principle, by assuming that very intense starbursts have occurred at high redshift , 
it would then be possible to explain the presence of metals in very underdense 
regions of the IGM even at redshift z$\sim 3-5$, i.e. when the universe was only $\sim 1.5$ Gyr old. 
We want to stress that these conclusions have been reached on the basis of pure order-of-magnitude arguments, and that 
only detailed numerical simulations following in detail the transport of metals in the IGM could give more 
precise estimates of the quantities described above. It would be very interesting to probe whether, 
with the prescriptions adopted above, in particular star formation rates of the order of $\ge 400 M_{\odot}/yr$, 
speeding up the ejecta at very high velocities, 
such simulations could be able to account for the presence of the metals on scales of $\sim 1 Mpc$ or more. 

\section{Conclusions} 
By means of chemo-photometric models for spheroidal  and dwarf irregular galaxies,  
we have carried out a detailed study of the history of 
element production in the universe. 
According to these models, the galaxy densities are fixed and are normalized to the 
values of the optical luminosity function in the local universe (Marzke et al. 1998). 
This is equivalent to assume that galaxies evolve in luminosity and not in number. 
 In our picture, 
the metal enrichment of the IGM is mainly due to galactic spheroids (i.e. ellipticals and spiral bulges) and, to a lesser extent, 
to dwarf irregular galaxies. We did not consider galactic winds in spiral discs. 
Our main results can be summarized as follows.\\
i) The enrichment of the IGM has proceeded in different manners for elements produced 
on different timescales.  In spheroids, 
$\alpha$-elements (O, Mg) are produced during the starburst phase, 
whereas for elements synthesized by low and intermediate mass stars (C, N, Fe)  
the bulk of the production 
occurs well after the starburst. The behaviour of Fe, 
produced mainly by type Ia SNe, is similar to the one of the elements produced by 
low and intermediate mass stars. 
Zn is also perhaps produced by type Ia SNe, but  a non-negligible 
contribution comes by massive stars (Fran\c cois et al. 2004). 
Dwarf irregular galaxies, characterized by a prolonged star formation history, in general have a different metal share than spheroids, 
with a low fraction locked up in stars 
(less than $\sim 10 \%$), $\sim 20 \%$ in the ISM and the highest fraction ( $\sim 70 \%$) ejected into the IGM. \\
ii) The bulk of the IGM enrichment is due to spheroids, with dwarf irregular galaxies playing a negligible role. 
Our prediction can reproduce the  [O/H]  observed in the IGM by various authors, 
as well as the trend of [O/H] with redshift.  
On the other hand, 
the predicted [C/H] are larger than observed, 
especially at high-redshift, in 2 out of 3 cases. 
A possible explanation could be that in regions of the universe with density close to the cosmic mean, where the vast majority of the baryons lie, 
  $C_{IV}$ loses its sensitivity as a metallicity tracer, since C has been ionized to higher levels, whereas $O_{VI}$ is a much more sensitive metallicity probe. 
Other possible reasons for the discrepancies between predictions and data for both C and O could be the 
fact that the IGM is not well mixed, or  
the data could be biased and representative of the average cosmic mean, or a  
a high fraction of the metals could not be observable, residing  at higher temperature in the intra-cluster hot gas, 
or warm gas.\\
iii) The study of the abundance ratios may be  useful to constrain the IMF of the first stars in galaxies.  
We have tried to fit the [Si/C] and [O/C] ratio observed in the local and 
high-redshift IGM by assuming a standard Salpeter IMF and a modified Salpeter 
flattened below  various possible mass values. 
 Our predictions underestimate the [O/C] and [Si/C] 
observed in the IGM at redshift larger than 2. To reduce this discrepancy, 
the adoption of a top-heavy IMF is not a viable solution, 
since it would overestimate the 
metal budget in the local universe (CM04). 
The explanation resides perhaps in  
inhomogeneities and local effects, which could be investigated by means of numerical simulations studying single elements and 
taking into account the stellar lifetimes. \\
iv) On the basis of simple but realistic physical arguments we have shown that, 
in order to disperse the metals up to distance scales  of $\sim 1 Mpc$ in less than $1 Gyr$, outflow velocities of  
$v_{W} \sim 1500 Km/s$  for the galactic winds are required. These velocities are  
of the same order of magnitude of the values observed in luminous infrared galaxies, 
experiencing the most intense starbursts in the universe. If present also at high redshift, as many recent observations seem to suggest 
(e.g. P\'erez-Gonz\'alez et al. 2005), 
these objects could be responsible for the enrichment of the lowest density regions of the IGM.\\
We stress that at the present time, the observations of the metal abundance in the IGM have been obtained for a few elements 
(mainly C and O). By improving the number of detections for these elements and by deriving abundances for other ones, it will 
be possible to shed more light on the sources of the IGM metals and on the mechanisms governing the IGM enrichment.  

\section*{Acknowledgments}
FC thanks Antonio Pipino for many helpful suggestions, 
Cristina Chiappini, Bob Carswell, Max Pettini and Valentina D'Odorico, Giovanni Vladilo for interesting discussions, and 
the hospitality of the Institute of Astronomy in Cambridge, where 
part of this work was carried out. 
The authors acknowledge funds from MIUR, COFIN 2003, prot. N. 2003028039.

\label{lastpage}

\end{document}